\documentclass[12pt,preprint]{aastex}

\usepackage[intlimits]{amsmath}
\usepackage{times}

\newcommand{\xt}{\tilde{x}}
\newcommand{\yt}{\tilde{y}}
\newcommand{\rt}{\tilde{r}}

\newcommand{\Dt}{\tilde{D}}
\newcommand{\htild}{\tilde{h}}

\newcommand{\Rt}{\tilde{R}}
\newcommand{\Eout}{E_{\text{\scriptsize out}}}

\shorttitle{Diffraction Analysis of Pupil Mapping}
\shortauthors{Vanderbei et.al.}

\begin{document}

\title{Diffraction Analysis of 2-D Pupil Mapping for High-Contrast Imaging}

\author{Robert J. Vanderbei}
\affil{Operations Research and Financial Engineering, Princeton University}
\email{rvdb@princeton.edu}

%

\begin{abstract} 
Pupil-mapping is a technique whereby a uniformly-illuminated
input pupil, such as from starlight, can be mapped into a non-uniformly
illuminated exit pupil, such that the image formed from this pupil will have
suppressed sidelobes, many orders of magnitude weaker than classical Airy ring
intensities.  Pupil mapping is therefore a candidate technique for
coronagraphic imaging of extrasolar planets around nearby stars.  
Unlike most other high-contrast imaging techniques,
pupil mapping is lossless and preserves the full 
angular resolution of the collecting telescope.  So, it could 
possibly give the highest signal-to-noise ratio of any proposed 
single-telescope system for detecting extrasolar planets.
Prior analyses based on pupil-to-pupil ray-tracing indicate that 
a planet fainter than $10^{-10}$ times its parent star, and as close as 
about $2 \lambda/D$, should be detectable.  
In this paper, we describe the results of careful diffraction analysis of
pupil mapping systems.  These results reveal a serious unresolved issue.
Namely, high-contrast pupil mappings distribute light from very near the edge
of the first pupil to a broad area of the second pupil and this dramatically
amplifies diffraction-based edge effects resulting in a limiting
attainable contrast of about $10^{-5}$.  We hope that by identifying this
problem others will provide a solution.
\end{abstract}


\section{Introduction} \label{sec:intro}

Pupil mapping for the high-contrast imaging required by the problem of finding
and imaging extra-solar terrestrial planets was first proposed by
\cite{Guy03}.  This idea has generated lots of excitement since it uses
$100\%$ of the available light and exploits the full resolution of the optical
system.  Preliminary laboratory results were presented in \citet{GGRSO04}.

In \citet{TV03} and \cite{VT04}, 
we studied pupil mapping as a method for generating
arbitrary pupil apodizations and, in particular, apodizations that provide the
ultra-high contrast needed for terrestrial planet finding.
By pupil mapping we mean a system of two lenses, or mirrors, that take a flat
input field at the entrance pupil and produce an output field that
is amplitude modified but still flat in phase (at least for on-axis sources).

\begin{figure}
\begin{center}
\includegraphics[width=2.5in]{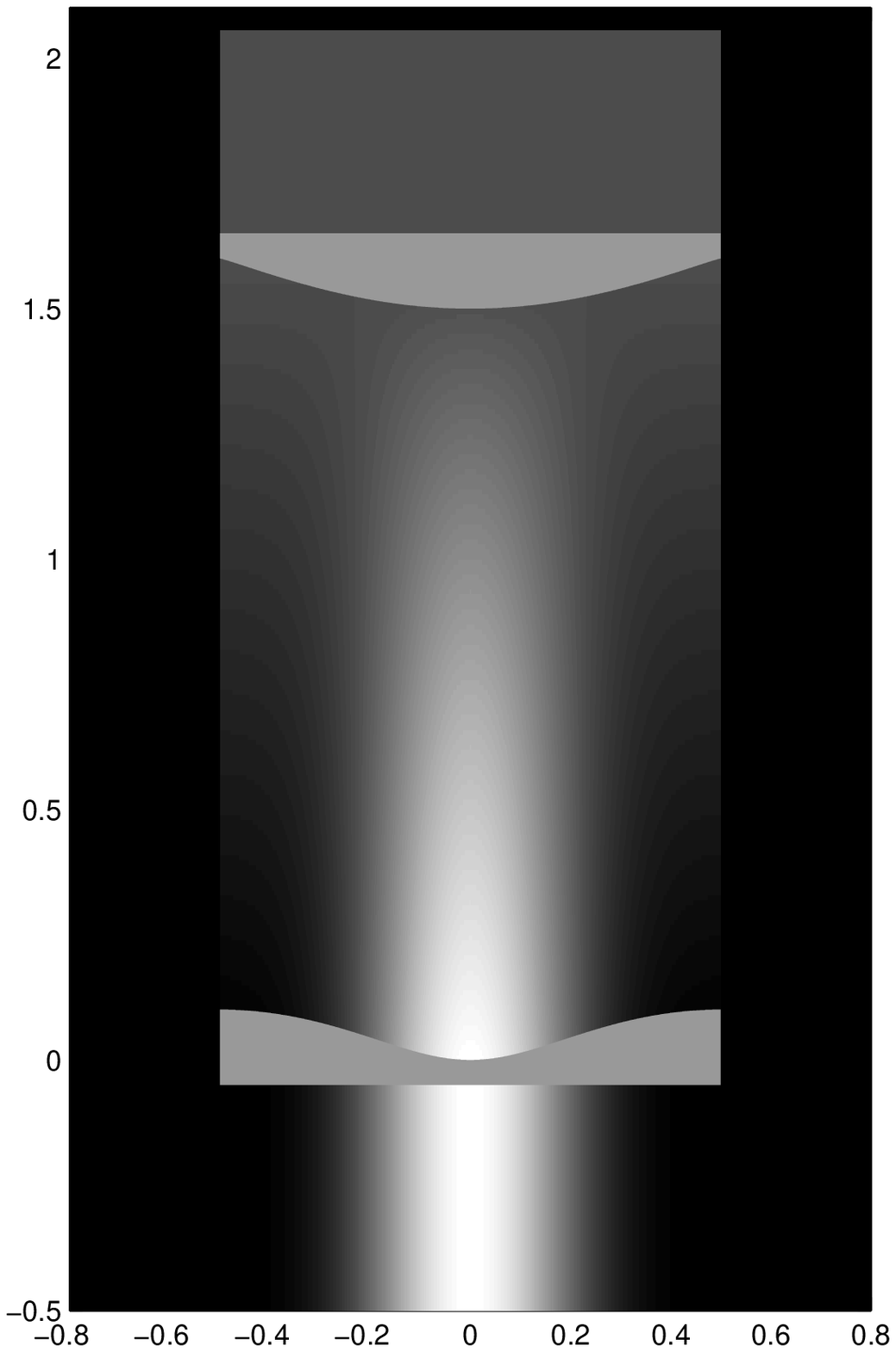}
\includegraphics[width=2.8in]{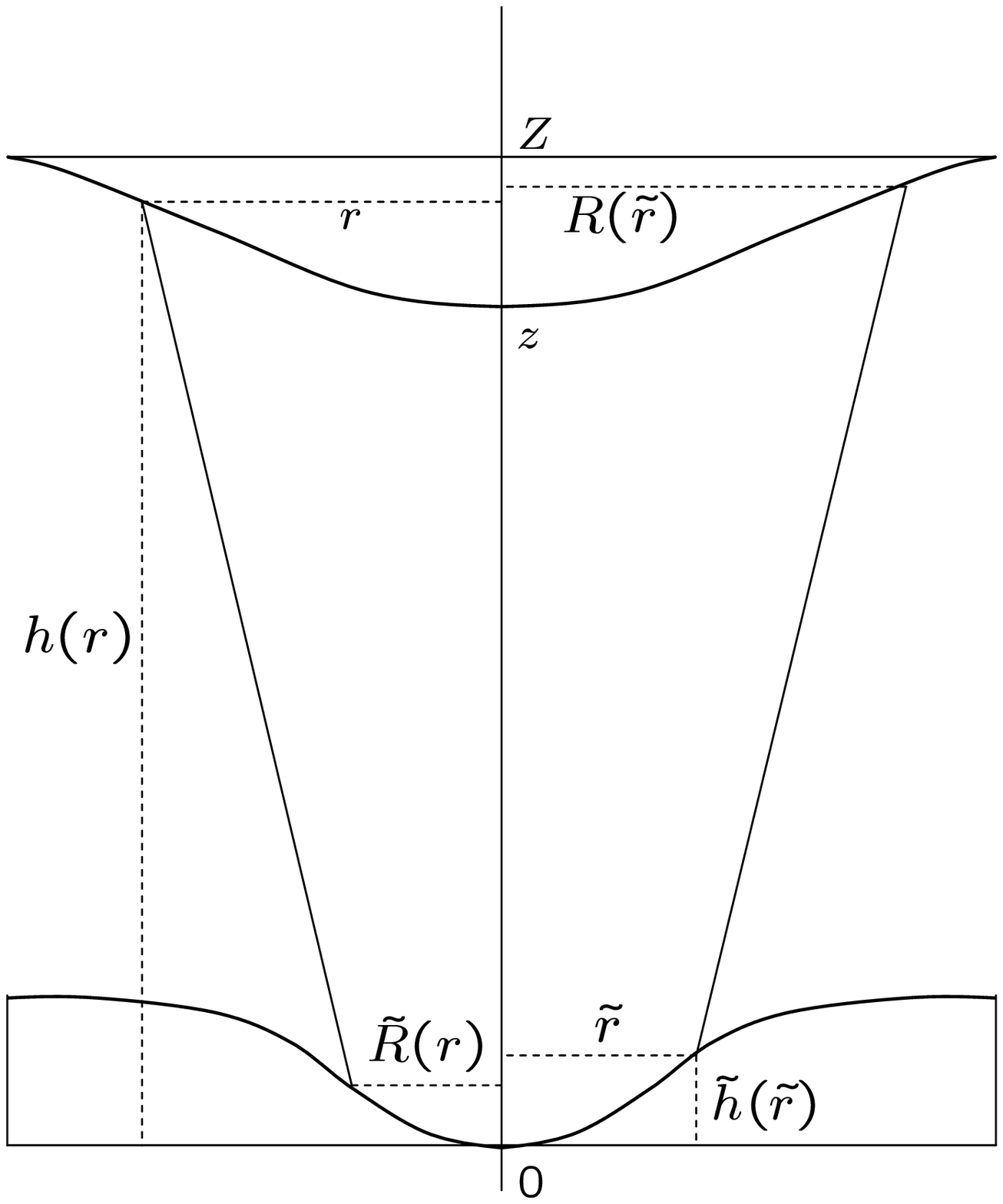}
\end{center}
\caption{Pupil mapping via a pair of properly figured lenses.}
\label{fig:1}
\end{figure}

Pupil mapping is easiest to describe in terms of ray-optics.  
An on-axis ray entering the
first pupil at radius $r$ from the center is to be mapped to radius 
$\rt = \Rt(r)$ at the exit pupil.
Optical elements at the two pupils ensure that the exit ray is 
parallel to the entering ray.  
The function $\Rt(r)$ is assumed to be
positive and increasing or, sometimes, negative and decreasing.  In either
case, the function has an inverse that allows us to recapture $r$ as a
function of $\rt$: $r = R(\rt)$.  
The purpose of pupil mapping is to create nontrivial amplitude apodizations.
An amplitude apodization function $A(\rt)$ specifies the ratio between the
output amplitude at $\rt$ to the input amplitude at $r$ (although we typically
assume the input amplitude is a constant).
We showed in \citet{VT04} that for any
amplitude apodization function $A(\rt)$ there is a pupil mapping
function $R(\rt)$ that achieves this amplitude profile.  Specifically, the
pupil mapping is given by
\begin{equation}\label{1}
    R(\rt) = \pm \sqrt{\int_0^{\rt} 2 A^2(s) s ds} .
\end{equation}
Furthermore, if we consider the case of a pair of lenses that are plano on
their outward-facing surfaces (as shown in Figure \ref{fig:1}), 
then the inward-facing surface profiles, $h(r)$ and $\htild(\rt)$, that
are required to obtain the desired pupil mapping are given by the solutions to
the following ordinary differential equations:
\begin{equation}\label{2}
    \frac{\partial h}{\partial r}(r) 
    = \frac{r-\Rt(r)}{ \sqrt{Q_0^2 + (n^2-1)(r-\Rt(r))^2} },
    \qquad
    h(0) = z,
\end{equation}
and
\begin{equation}\label{3}
    \frac{\partial \htild}{\partial \rt}(\rt) 
    = \frac{R(\rt)-\rt}{ \sqrt{\phantom{\tilde{|}}Q_0^2 + (n^2-1)(R(\rt)-\rt)^2} },
    \qquad
    \htild(0) = 0.
\end{equation}
Here, $n$ is the refractive index and 
$Q_0$ is a constant determined by the distance $z$ separating the centers
($r=0$, $\rt = 0$) of the two lenses:
$Q_0 = -(n-1)z$.

Let $S(r,\rt)$ denote the distance between a point on the
first lens surface $r$ units from the center and the corresponding point on
the second lens surface $\rt$ units from its center.  Up to an additive
constant, the optical path length of a ray that exits at radius 
$\rt$ after entering at radius $r = R(\rt)$ is given by
\begin{equation}\label{4}
    Q_0(\rt) = S(R(\rt),\rt) + n(\htild(\rt)-h(R(\rt))) .
\end{equation}
In \citet{VT04}, we showed that, for an on-axis source,
$Q_0(\rt)$ is constant and equal to $Q_0$.

\section{High-Contrast Apodization} \label{sec:apod}

\begin{figure}
\begin{center}
\text{\includegraphics[width=3.0in]{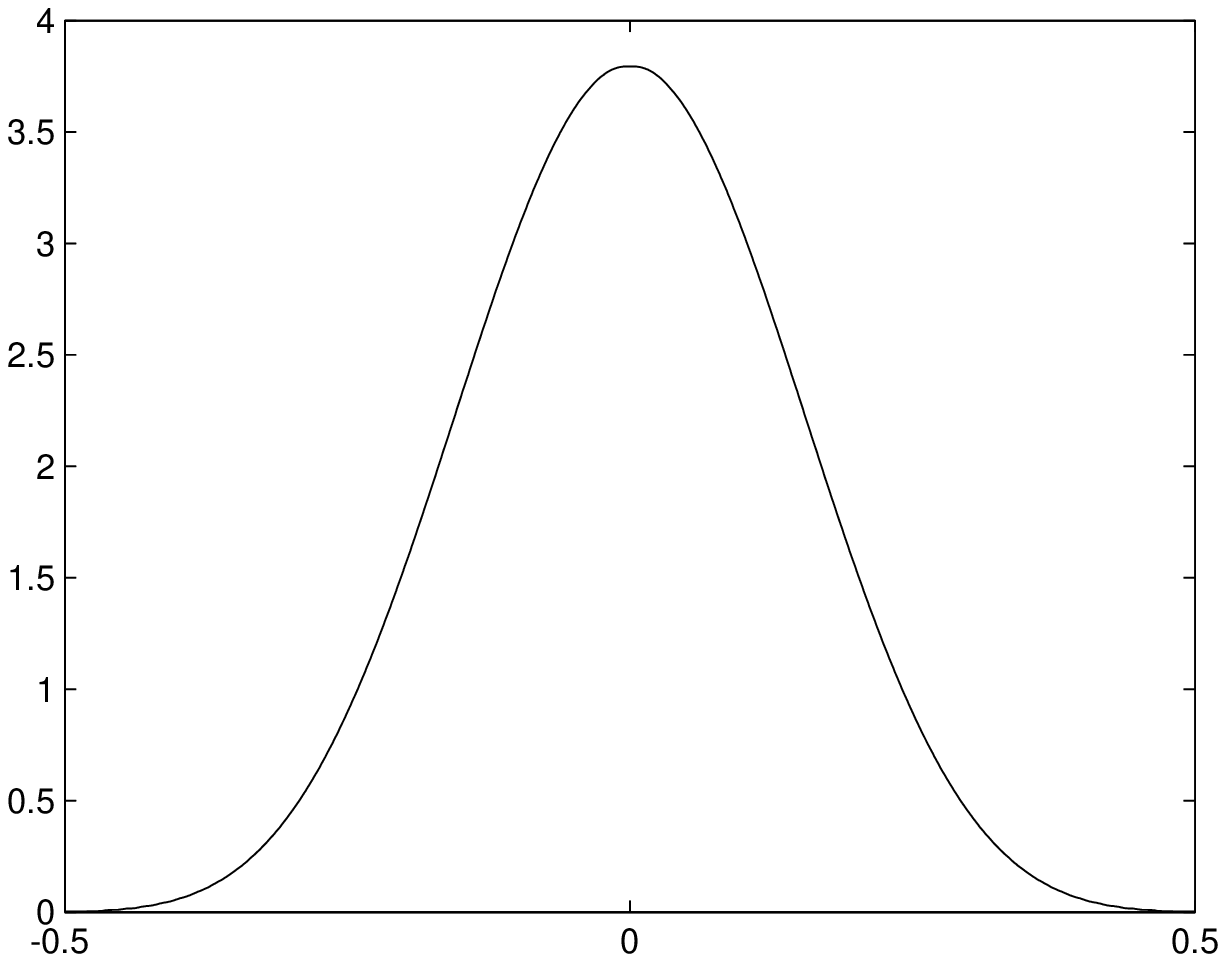}}
\text{\includegraphics[width=3.0in]{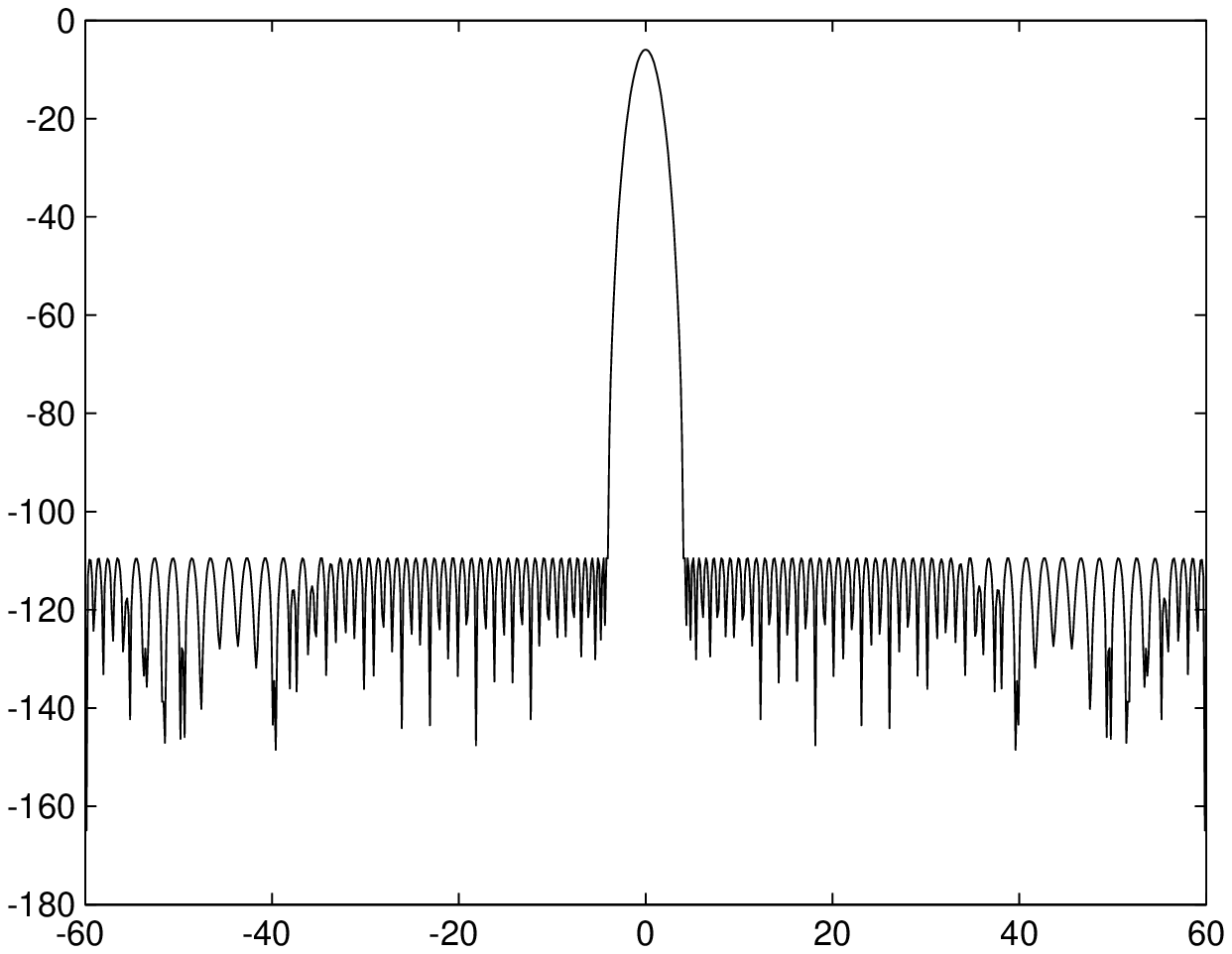}}
\end{center}
\caption{{\em Left.} 
An apodization providing contrast of $10^{-10}$ at tight inner working angles.
{\em Right.} The corresponding on-axis point spread function.
}
\label{fig:2}
\end{figure}

If we assume that an apodized beam with amplitude apodization profile $A(\rt)$
such as one obtains as the output of a pupil mapping system
is passed into an ideal imaging system with focal length $f$, the electric
field $E(\rho)$ at the image plane is given by the Fourier transform 
of $A(\rt)$:
\begin{equation}\label{5}
    E(\xi,\eta) 
    =  
    \frac{E_0}{\lambda f} \iint e^{2 \pi i \frac{\xt\xi + \yt\eta}{\lambda f}} 
          A(\sqrt{\xt^2+\yt^2}) d\yt d\xt.
\end{equation}
Here, $E_0$ is the input amplitude which, unless otherwise noted, 
we take to be unity.
Since the optics are azimuthally symmetric, it is convenient to use polar
coordinates.  The apodization function $A$ is a function of 
$\rt = \sqrt{\xt^2+\yt^2}$ and the image-plane electric field depends only on
image-plane radius $\rho = \sqrt{\xi^2 + \eta^2}$:
\begin{eqnarray}
    E(\rho)
    & = & 
    \frac{1}{\lambda f}
    \iint e^{2 \pi i \frac{\rt \rho}{\lambda f} \cos(\theta - \phi)} 
                A(\rt) \rt d\theta d\rt \label{6} \\
    & = & 
    \frac{2 \pi}{\lambda f}
    \int J_0\left(2 \pi \frac{\rt \rho}{\lambda f}\right) 
               A(\rt) \rt d\rt . \label{7}
\end{eqnarray}
The point-spread function (PSF) is the square of the electric field:
\begin{equation}\label{8}
    \mbox{Psf}(\rho) = |E(\rho)|^2 .
\end{equation}
For the purpose of terrestrial planet finding, it is important to construct an
apodization for which the PSF at small nonzero angles is ten orders of
magnitude reduced from its value at zero.  Figure \ref{fig:2} shows one such
apodization function.  In \citet{VSK03},
we explain how these apodization functions are computed.

\section{Huygens Wavelets} \label{sec:huygens}

We have designed the pupil mapping system using simple ray optics but we have
relied on diffraction theory to ensure that the apodization provides high
contrast.  This begs the question as to whether the desired high contrast will
remain after a diffraction analysis of the entire system including the
two-lens pupil mapping system or
will diffraction effects in the pupil mapping system itself create ``errors''
that are great enough to destroy the high-contrast that we seek.  To
answer this question, we need to do a diffraction analysis of
the pupil mapping system itself.

If we assume that a flat, on-axis, electric field arrives at the entrance 
pupil, then the electric field at a particular point of the exit pupil 
can be well-approximated by superimposing the phase-shifted waves from each
point across the entrance pupil (this is the well-known Huygens-Fresnel
principle---see, e.g., Section 8.2 in \cite{BW99}).  That is,
\begin{equation}\label{9}
  \Eout(\xt, \yt)
  =
  \iint 
       \frac{1}{\lambda Q(\xt,\yt,x,y)}
       e^{ 2 \pi i Q(\xt,\yt,x,y) /\lambda } 
       dydx,
\end{equation}
where
\begin{equation}\label{20}
       Q(\xt,\yt,x,y) =
       \sqrt{(x-\xt)^2+(y-\yt)^2+(h(r)-\htild(\rt))^2} 
       + n(Z - h(r) + \htild(\rt))
\end{equation}
is the optical path length, $Z$ is the distance between the plano lens
surfaces (i.e., a constant slightly larger than $z$),
and where, of course, we have used $r$ and $\rt$ as shorthands for the radii in
the entrance and exit pupils, respectively.  As before, it is convenient to 
work in polar coordinates:
\begin{equation}\label{10}
  \Eout(\rt)
  =
  \iint 
	\frac{1}{\lambda Q(\rt,r,\theta)}
        e^{ 2 \pi i Q(\rt,r,\theta) /\lambda} 
	r d\theta dr,
\end{equation}
where
\begin{equation}\label{21}
	Q(\rt,r,\theta) =
	\sqrt{r^2-2r\rt\cos\theta+\rt^2+(h(r)-\htild(\rt))^2} 
        + n(Z - h(r) + \htild(\rt)) .
\end{equation}
For numerical tractability, it is essential to make approximations so that 
the integral over $\theta$ can be carried out analytically.
To this end, we need to make an appropriate approximation to the square root
term:
\begin{equation}\label{13}
     S = \sqrt{r^2-2r\rt\cos\theta+\rt^2+(h(r)-\htild(\rt))^2} .
\end{equation}

\section{Fresnel Propagation} \label{sec:fresnel}

In this section we consider approximations that lead to the so-called
{\em Fresnel propagation formula}.

If we assume that the lens separation is fairly large, then the $(h-\htild)^2$
term dominates the rest and so we can use the first two terms of a Taylor 
series approximation (i.e., for $u$ small relative to $a$,
$\sqrt{u^2+a^2} \approx a + u^2/2a$) to get the following {\em large
separation} approximation:
\begin{equation}\label{11}
	S
	\approx
	(h(r)-\htild(\rt)) +
	\frac{r^2 - 2r \rt \cos\theta + \rt^2}{2(h(r)-\htild(\rt))} .
\end{equation}

If we assume further that the lenses are thin (i.e., that $n$ is large),
then $h-\htild$ in the denominators can be approximated simply by $z$:
\begin{equation}\label{12}
    S \approx (h(r)-\htild(\rt)) + \frac{r^2 - 2r \rt \cos\theta + \rt^2}{2z} .
\end{equation}
This is called the {\em thin lens approximation}.

Combining the large lens separation approximation with the thin lens
approximation, we get
\begin{equation}
  \Eout(\rt)
  = \iint 
	\frac{1}{\lambda Q_1(\rt,r,\theta)}
        e^{ 2 \pi i Q_1(\rt,r,\theta) /\lambda }
	r d\theta dr\label{14} 
\end{equation}
where
\begin{equation}\label{23}
	Q_1(\rt,r,\theta)
	=
	\frac{r^2 - 2r \rt \cos\theta + \rt^2}{2z} 
        + Z + (n-1)(Z - h(r) + \htild(\rt)).
\end{equation}
Finally, we simplify the reciprocal of $Q_1$ by noting that
the $Z$ term dominates the other terms (i.e., for $u$ small relative to $a$, 
$1/(u+a) \approx 1/a$) and so we get that:
\begin{equation}\label{24}
	\frac{1}{Q_1(\rt,r,\theta)} \approx \frac{1}{Z} .
\end{equation}
This last approximation is called the {\em paraxial} approximation.

Combining all three approximations, 
we now arrive at the {\em standard Fresnel approximation}:
\begin{equation}\label{15}
  \Eout(\rt)
  = \frac{2 \pi}{\lambda Z}
        e^{\pi i \frac{\rt^2}{z\lambda} 
	 + 2 \pi i \frac{(n-1)\htild(\rt)}{\lambda}} 
	\int
        e^{\pi i \frac{r^2}{z\lambda}  
	 - 2 \pi i \frac{(n-1)h(r)}{\lambda}} 
	J_0( 2 \pi r \rt / z \lambda )
	r dr.
\end{equation}

While the standard Fresnel approximation works very well in most
conventional situations, it turns out (as well shall show) 
to be too crude of an approximation for high-contrast pupil mapping.
It is inadequate because it does not honor the constancy of the optical 
path length $Q(\rt,r,\theta)$ along the rays of ray-optics.  That is, 
the fact that $Q(\rt,R(\rt),0)$ is constant
has been lost in the approximations.
We should have used the ray-tracing optical path length as the ``large
quantity'' in our large-lens-separation approximation instead of the simpler
difference $h(r)-\htild(\rt)$.  But, this seemingly simple adjustment quickly
gets tedious and so we prefer to take a completely different (and simpler)
approach, which is described in the next section.

\section{An Alternative to Fresnel} \label{sec:alternative}

As we have just explained and shall demonstrate later, the standard Fresnel 
approximation does not produce good results for high-contrast
pupil mapping computations.
In this section, we present an alternative approximation that is slightly more
computationally demanding but is much closer to a direct calculation of 
the true Huygens wavelet propagation.

As with Fresnel, we approximate the $1/Q(\rt,r,\theta)$ 
amplitude-reduction
factor in \eqref{10} by the constant $1/Z$ (the {\em paraxial approximation}).
The $Q(\rt,r,\theta)$ appearing in the exponential must, on the other hand,
be treated with care.
Recall that $Q(\rt,R(\rt),0)$ is a constant.  Since constant phase shifts are
immaterial, we can subtract it from $Q(\rt,r,\theta)$ in \eqref{10} to get
\begin{equation} \label{140}
  \Eout(\rt)
  \approx
  \frac{1}{\lambda Z}
  \iint 
        e^{ 2 \pi i (Q(\rt,r,\theta)-Q(\rt,R(\rt),0)) /\lambda} 
	r d\theta dr.
\end{equation}
Next, we write the difference in $Q$'s as follows:
\begin{eqnarray} 
    Q(\rt,r,\theta)-Q(\rt,R(\rt),0) 
    & = &
    S(\rt,r,\theta)-S(\rt,R(\rt),0)
    + n(h(R(\rt)) - h(r)) \label{141} \\
    & = &
    \frac{S^2(\rt,r,\theta)-S^2(\rt,R(\rt),0)}{S(\rt,r,\theta)+S(\rt,R(\rt),0)}
    + n(h(R(\rt)) - h(r)) \label{142} 
\end{eqnarray}
and then we expand out the numerator and cancel big terms that can be
subtracted one from another to get
\begin{eqnarray} 
    S^2(\rt,r,\theta)-S^2(\rt,R(\rt),0)
    & = & 
    (r - R(\rt)) (r + R(\rt)) - 2\rt\left(r \cos \theta - R(\rt)\right) \notag\\
    && \qquad + 
    \left(h(r)-h(R(\rt))\right) \left(h(r)+h(R(\rt))-2\htild(\rt)\right) .
\end{eqnarray}
When $r = R(\rt)$ and $\theta = 0$,
the right-hand side clearly vanishes as it should.
Furthermore, for $r$ close to $R(\rt)$ and $\theta$ close to zero,
the right-hand side gives an accurate formula for computing the deviation from
zero.  That is to say, the right-hand side is easy to program in such a manner
as to avoid subtracting one large number from another, which is always the
biggest danger in numerical computation.

So far, everything is exact (except for the paraxial approximation).
The only further approximation we make is to replace 
$S(\rt,r,\theta)$ in the denominator of \eqref{142} with $S(\rt,R(\rt),0)$
so that the denominator becomes just $2S(\rt,R(\rt),0)$.
Putting this altogether and replacing the integral on $\theta$ with the
appropriate Bessel function, we get a new approximation, which we refer to as
the {\em Huygens} approximation:
\begin{eqnarray} 
  \Eout(\rt)
  & \approx &
  \frac{2 \pi}{\lambda Z}
  \int 
    e^{ 2 \pi i \left(
      \frac{
        (r - R(\rt)) (r + R(\rt)) + 2\rt R(\rt)
        + 
        \left(h(r)-h(R(\rt))\right) \left(h(r)+h(R(\rt))-2\htild(\rt)\right) 
      }{2S(\rt,R(\rt),0)}
      + n(h(R(\rt)) - h(r)) 
    \right) /\lambda} \notag \\
    && \qquad \qquad \times
    J_0\left(2 \pi \rt r / \lambda S(\rt,R(\rt),0)\right)
    r dr.\label{144}
\end{eqnarray}

\section{Sanity Checks} \label{sec:examples}

In this section we consider a number of examples.

\subsection{Flat glass windows ($A \equiv 1$)}

\begin{figure}
\begin{center}
\text{\includegraphics[width=6.5in]{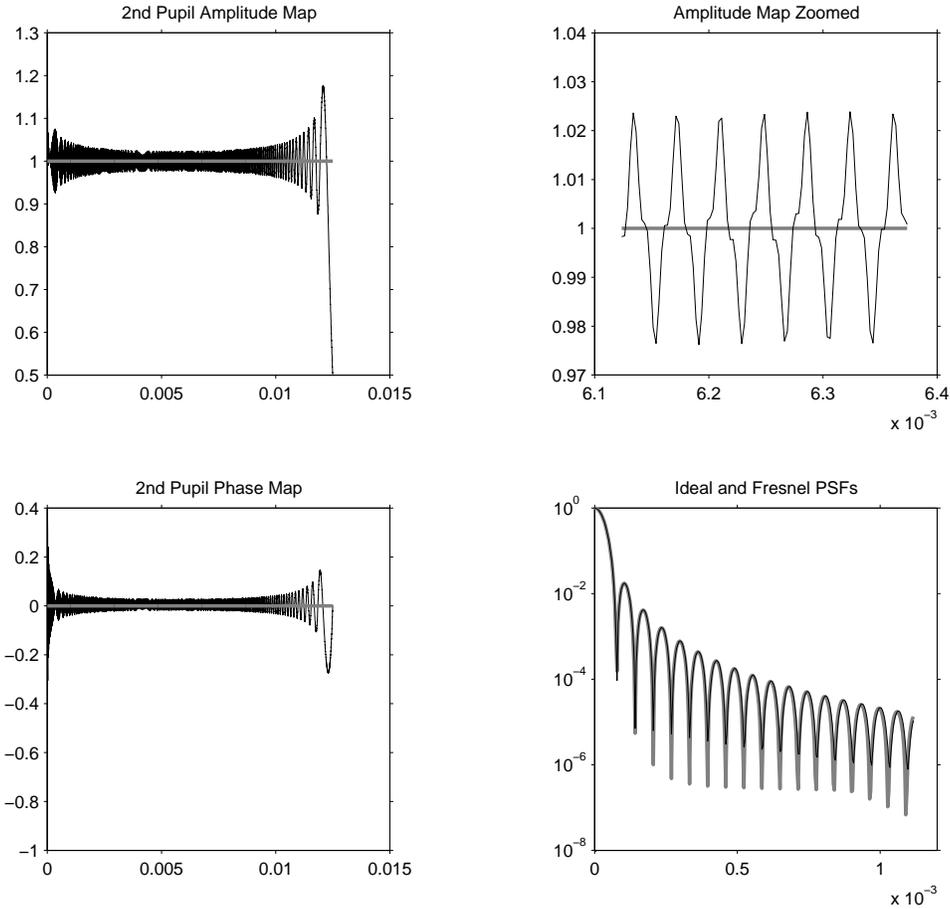}}
\end{center}
\caption{Fresnel analysis of a pupil mapping system consisting of two flat
pieces of $n=1.5$ glass in circular $D = 25$-mm 
apertures separated by $z = 15D$.  The wavelength is $\lambda = 632.8$-nm.
{\em Upper-left} plot shows in gray the target amplitude apodization profile and 
in black the amplitude profile computed using standard Fresnel propagation.
{\em Upper-right} plot shows a zoomed in section of the amplitude profiles.
{\em Lower-left} plot shows in gray the computed optical path length $Q_0(\rt)$
and in black the phase map computed using Fresnel propagation.  
{\em Lower-right} plot shows in gray the PSF associated with the ideal
amplitude apodization and in black the PSF computed by Fresnel propagation.
These results should match
textbook examples for simple open circular apertures---and they do.
}
\label{fig:3}
\end{figure}

We begin with the simplest example in which the apodization function is
identically equal to one.  Taking the positive root in \eqref{1}, we get that
$R(\rt) = \rt$.  That is, the ray-optic design is for the light to go straight
through the system.  The inverse map $\Rt(r)$ is also trivial: $\Rt(r) = r$.
Hence, the right-hand sides in the differential equations \eqref{2} and 
\eqref{3} vanish and the lens figures become flat: $h(r) \equiv z$ and
$\htild(\rt) \equiv 0$.  In this case, $Q_1(\rt,R(\rt),0)$ is a constant 
(independent of $\rt$) and so the Fresnel approximation is a good one.
The Fresnel results, shown in Figure \ref{fig:3} should match
textbook examples for simple open circular apertures---and they do.

\subsection{A Galilean Telescope ($A \equiv a > 1$)}

\begin{figure}
\begin{center}
\text{\includegraphics[width=6.5in]{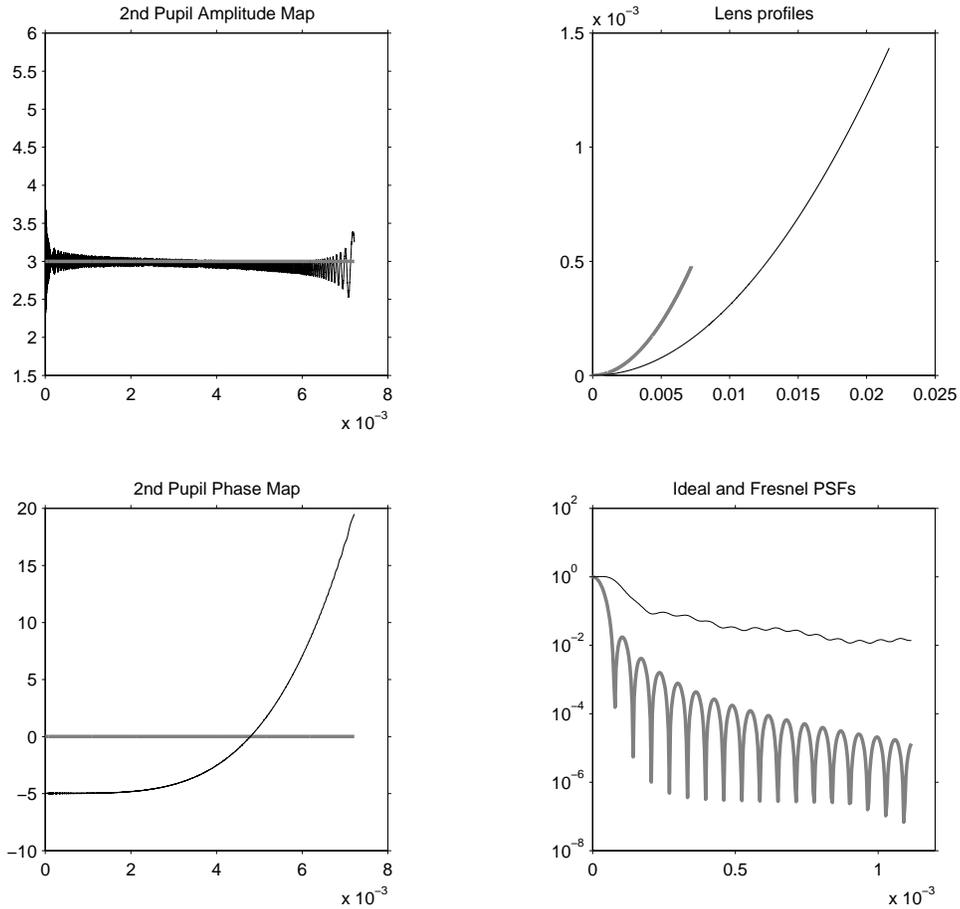}}
\end{center}
\caption{Same as in Figure \ref{fig:3} but with the apodization function
replaced with a constant value of $A\equiv3$.  In this example, both
lenses are hyperbolic as shown in the upper-right plot.  
The lens profiles $h$ and $\htild$ were computed using a 
$5,000$ point
discretization and the Fresnel propagation (equation \eqref{15}) 
was carried out also with a $5,000$ point discretization.
Note the large error in the phase map and the fact that
the computed PSF does not follow the usual Airy pattern.  
This is strong evidence that the standard Fresnel approximation is
too crude since real systems of this sort
exist and exhibit the expected Airy pattern.
}
\label{fig:4}
\end{figure}

In this case, we ask for a system in which the output pupil has been uniformly
amplitude intensified by a factor $a > 1$.  
If we choose the positive root in \eqref{1}, then we get a
Galilean-style refractor telescope consisting of a convex lens at the entrance
pupil and a concave lens as an eyepiece.  
Specifically, we get $R(\rt) = a \rt$ and $\Rt(r) = r/a$.
From these it follows that if the aperture of the first lens is $D$, then the
aperture of the second is $\Dt = D/a$.
It is easy to compute the lens figures
\begin{eqnarray} 
    h(r) &=& z + \frac{\sqrt{Q_0^2+(n^2-1)(1-1/a)^2r^2}-|Q_0|}{(n^2-1)(1-1/a)}
	    \label{201} \\[0.2in]
    \htild(\rt) &=& \frac{\sqrt{Q_0^2+(n^2-1)(a-1)^2\rt^2}-|Q_0|}{(n^2-1)(a-1)}
	    \label{202}.
\end{eqnarray}
If the relative index of refraction $n$ is greater than $1$ (as in air-spaced
glass lenses), then these functions represent portions of a hyperbola.
If, on the other hand, $n<1$ (as in a glass medium between the two surfaces),
then the functions are ellipses.  Fresnel
results for $a=3$ and $n=1.5$ are shown in Figure \ref{fig:4}.
Note the large error in the phase map and the fact that
the computed PSF does not follow the usual Airy pattern.  
This is strong evidence that the Fresnel approximation is
too crude since real systems of this sort
exist and exhibit the expected Airy pattern.

\begin{figure}
\begin{center}
\text{\includegraphics[width=6.5in]{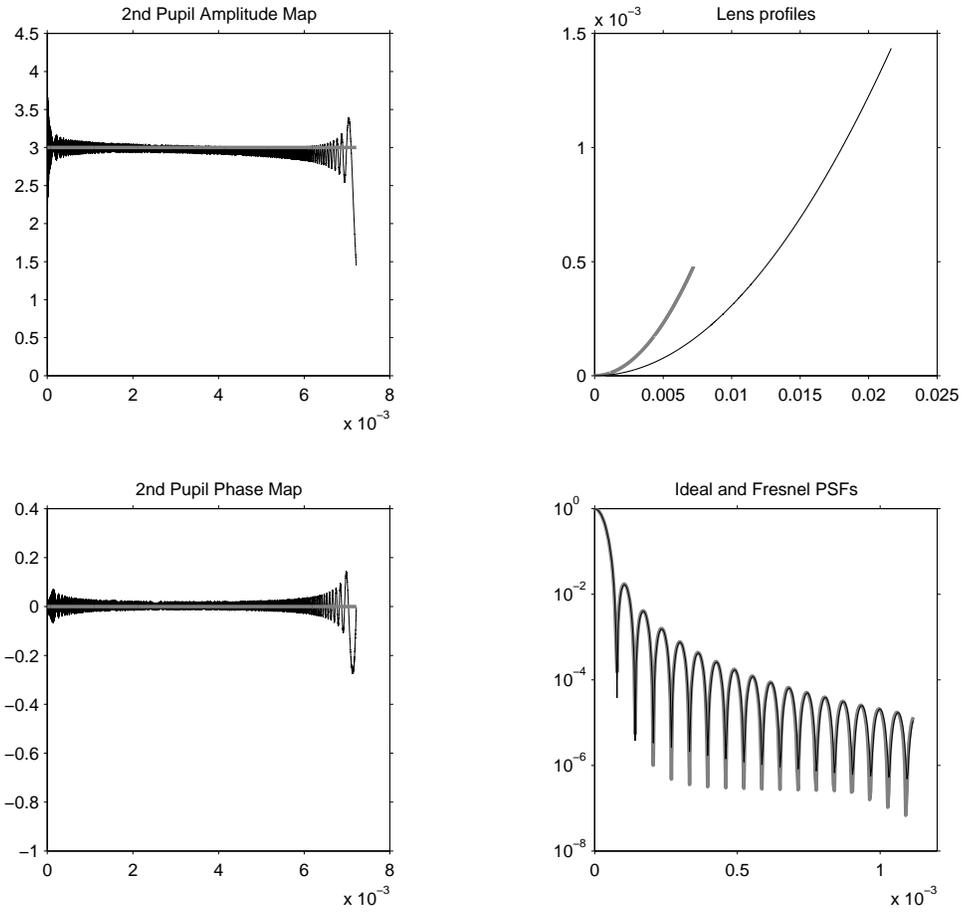}}
\end{center}
\caption{Same as in Figure \ref{fig:4} but computed using the Huygens
approximation.  
Note that the phase map, while not perfect, is now much flatter.
Also, the computed PSF is closely matches the expected Airy pattern.
}
\label{fig:5}
\end{figure}

In Figure \ref{fig:5} we consider the same system but use 
the Huygens approximation.  
Note that the phase map, while not perfect, is now much flatter.
Also, the computed PSF is closely matches the expected Airy pattern.


\subsection{An Ideal Lens} \label{sec:ideal}

If we let $a$ tend to infinity, we see that $\Dt$ tends to zero and the
system reduces to a convex lens focusing a collimated input beam to a point.
In this case, the second lens vanishes; only the first lens is of
interest.  Its equation is
\begin{equation} \label{203}
    h(r) = z + \frac{\sqrt{Q_0^2+(n^2-1)r^2}-|Q_0|}{(n^2-1)}.
\end{equation}
The plane where the second lens was is now the image plane.
To compute the electric field here, we put $\htild(\rt) \equiv 0$ and use
either the Fresnel or the Huygens approximation.  Since we would put
a detector at this plane, we can ignore any final phase corrections and both
approximations reduce to the same formula:
\begin{equation} \label{204}
  \Eout(\rt)
  = \frac{2 \pi}{\lambda Z}
	\int
        e^{\pi i \frac{r^2}{z\lambda}  
	 - 2 \pi i \frac{(n-1)h(r)}{\lambda}} 
	J_0( 2 \pi r \rt / z \lambda )
	r dr.
\end{equation}
Substituting \eqref{203} into \eqref{204} and dropping any unit complex
numbers that factor out of the integral, we get
\begin{equation} \label{205}
  \Eout(\rt)
  = \frac{2 \pi}{\lambda Z}
	\int
        e^{\pi i \frac{r^2}{z\lambda}  
	 - \frac{2 \pi i}{\lambda} 
	   \sqrt{\left(\frac{n-1}{n+1}\right)^2 z^2 + \frac{n-1}{n+1}r^2}} 
	J_0( 2 \pi r \rt / z \lambda )
	r dr.
\end{equation}
Of course, this formula is for a uniform collimated input beam.
If the input beam happens to be apodized by some upstream optical element,
then the expression becomes
\begin{equation} \label{207}
  \Eout(\rt)
  = \frac{2 \pi}{\lambda Z}
	\int
        e^{\pi i \frac{r^2}{z\lambda}  
	 - \frac{2 \pi i}{\lambda} 
	   \sqrt{\left(\frac{n-1}{n+1}\right)^2 z^2 + \frac{n-1}{n+1}r^2}} 
	J_0( 2 \pi r \rt / z \lambda )
	A(r)
	r dr
\end{equation}
where $A(r)$ denotes the apodization function.
This formula does not agree with the Fourier transform expression given earlier
by equation \eqref{5}.  However, if the square root is approximated by the
first two terms of its Taylor expansion, 
\begin{equation} \label{206}
    \sqrt{\left(\frac{n-1}{n+1}\right)^2 z^2 + \frac{n-1}{n+1}r^2} 
    = \frac{n-1}{n+1} z + \frac{r^2}{2 z} ,
\end{equation}
then \eqref{207} reduces to a Fourier transform as in
\eqref{5} (again dropping unit complex factors).

\begin{figure}
\begin{center}
\text{\includegraphics[width=6.5in]{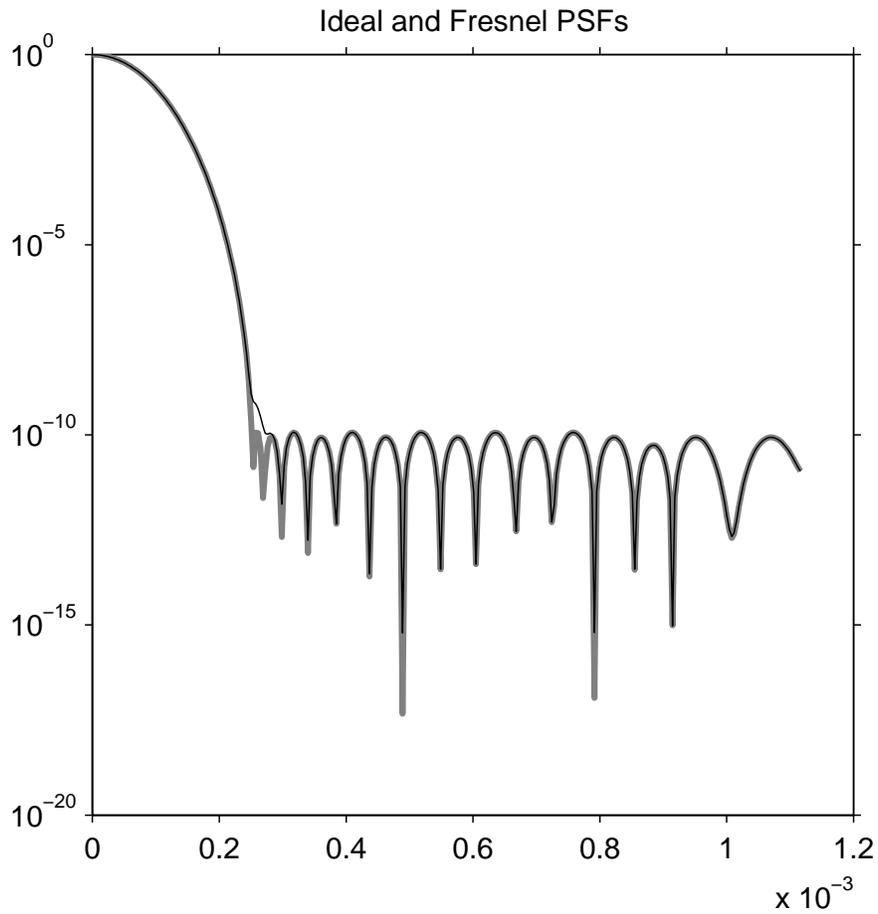}}
\end{center}
\caption{
The gray plot shows an 
ideal high-contrast PSF designed assuming that the focussing element 
is a parabolic mirror so that the Huygens image-plane electric field is 
a Fourier transform as in equation \eqref{5}.
The black line shows the PSF if one assumes that the 
focussing element is actually an ideal lens (having an elliptical profile)
as described in Section \ref{sec:ideal}.
The two curves are visually identical 
except in the neighborhood of the 1st and 2nd minima of the Fourier 
(gray) curve.
}
\label{fig:6}
\end{figure}

This raises an interesting question: if a high-contrast apodization is
designed based on the assumption that the focusing element behaves like a
Fourier transform (i.e., as in \eqref{5}), how well will the apodization work
if the true expression for the electric field is closer to the one given by
\eqref{207}?  The answer is shown in Figure \ref{fig:6}.  The PSF degradation
is very small.

\section{High-Contrast Apodization}

\begin{figure}
\begin{center}
\text{\includegraphics[width=6.5in]{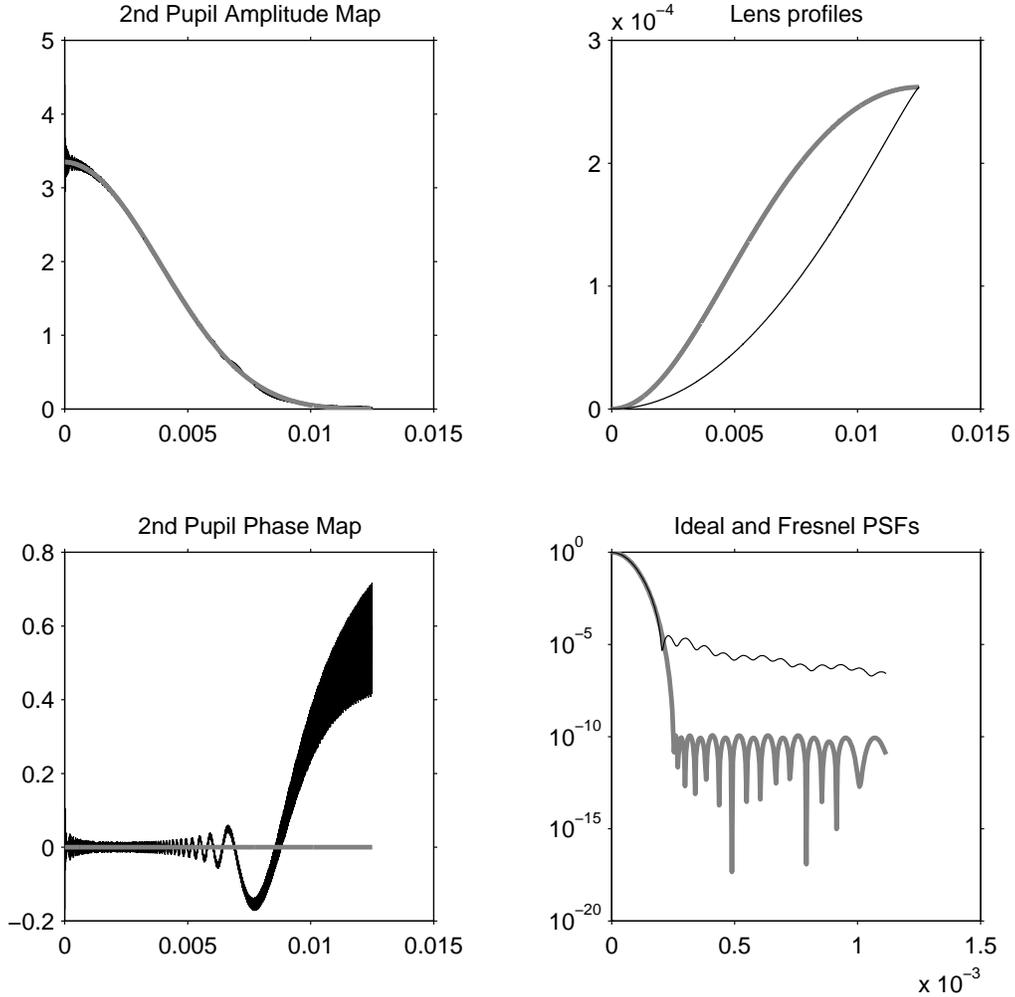}}
\end{center}
\caption{
Analysis of a pupil mapping system using the Huygens approximation with 
$z = 15D$ and $n=1.5$.
{\em Upper-left} plot shows in gray the target amplitude apodization profile and 
in black the amplitude profile computed using the Huygens approximation.
{\em Upper-right} plot shows the lens profiles, black for the first lens and
gray for the second.
The lens profiles $h$ and $\htild$ were computed using a $5,000$ point
discretization.
{\em Lower-left} plot shows in gray the computed optical path length $Q_0(\rt)$
and in black the phase map computed using Huygens propagation.  
The Huygens propagation was carried out with a $5,000$ point discretization.
{\em Lower-right} plot shows in gray the PSF computed as the square of the
Fourier transform of the ideal
amplitude apodization and in black the PSF computed using the Huygens
approximation.
}
\label{fig:7}
\end{figure}

The purpose of the examples discussed in the previous section was to convince
the reader that the Huygens approximation provides a reasonable
estimate of the electric field at the exit pupil of a pupil mapping system.
Assuming we were convincing, we now proceed to apply the Huygens
approximation to compute the electric field and focused PSF
of the pupil mapping system corresponding to the
apodization function shown in Figure \ref{fig:2}.  As before, we
assume a wavelength of
$0.6328\mu$ and lenses with aperture $D = 25$mm.  In Figure \ref{fig:7}, 
we show the
results for $z = 15D$ and a refraction index of $1.5$.  For these simulations
to be meaningful, it is critical that the integrals be represented by a sum
over a sufficiently refined partition.  The bigger the disparity between
wavelength $\lambda$ and aperture $D$, the more refined the partition needs to
be.  For the parameters we have chosen, a partition into $5000$ parts proved
to be adequate and is what we have chosen.  
The plot in the upper-left section of
the figure shows in gray the target amplitude apodization profile and 
in black the amplitude profile computed using the Huygens approximation (i.e.,
equation \eqref{144}).  The plot in the upper right shows the lens profiles.
The first lens is shown in black and the second in gray.
The plot in the
lower left shows in gray the computed optical path length $Q_0(\rt)$.  
If the numerical computation of the lens figures had been done with
insufficient precision, this curve would not be flat.  As we see, it is flat.
The lower left also shows in black the phase map as computed by the
Huygens approximation.
Note that here there are high frequency oscillations
everywhere and low frequency oscillation that has an amplitude that increases
as one moves out to the rim of the lens.
The lower-right plot shows in gray the PSF associated with the ideal
amplitude apodization and in black the PSF computed by Huygens propagation.

\begin{figure}
\begin{center}
\text{\includegraphics[width=6.5in]{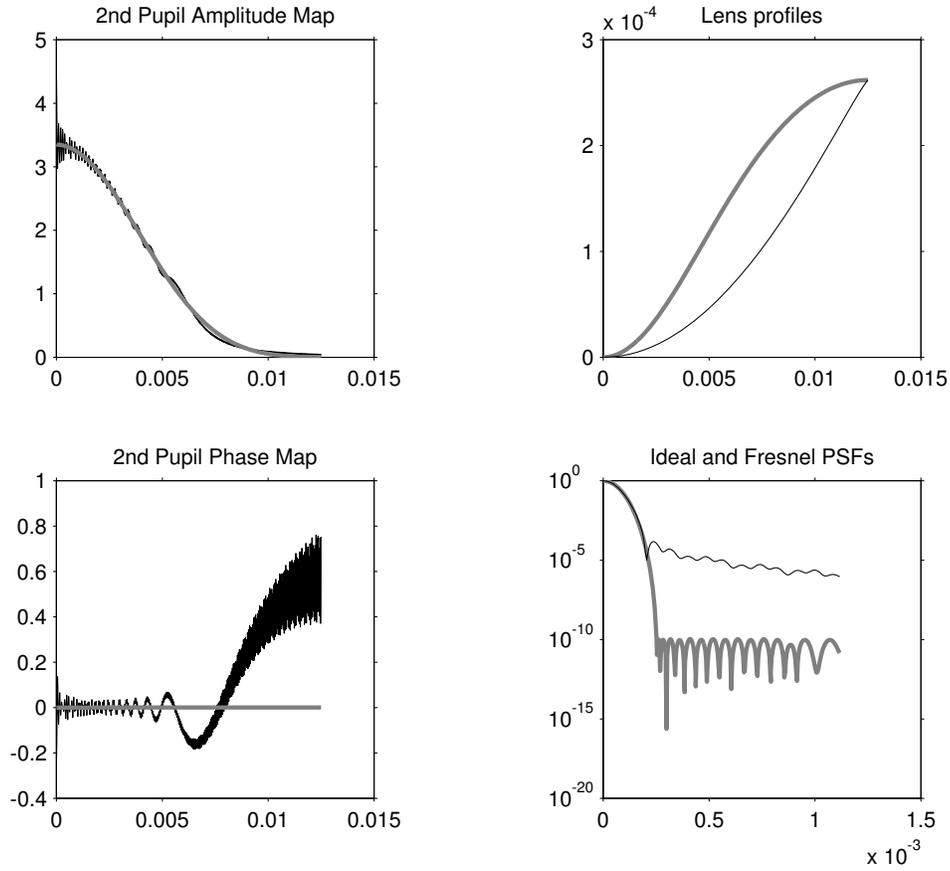}}
\end{center}
\caption{Same as in Figure \ref{fig:7} but computed using a brute force
computation of the Huygens integral \eqref{10}.  Here we used $1000$ $r$-values
and $1000$ $\theta$-values.  Consequently, we needed to increase the wavelength
by a factor $5$ to guarantee adequate sampling of the phase and amplitude
ripples.
}
\label{fig:8}
\end{figure}

The PSF in Figure \ref{fig:7} is disappointing.
It is important to determine whether this is real or is a result
of the approximations behind the Huygens propagation formula.
As a check, we did a brute force computation of the Huygens
integral \eqref{10}.  Because this integral is more difficult, we were forced
to use only 500 $r$-values and 500 $\theta$-values.  Hence, we had to
increase the wavelength by a factor of $10$.  With these changes, the
result is shown in Figure \ref{fig:8}.  It too shows the same amplitude and
phase oscillations.  This sanity check convinces us that these effects are
physical.  We need to consider changes to the physical setup that might
mollify these oscillations.  Such changes are considered next.

\begin{figure}
\begin{center}
\text{\includegraphics[width=6.5in]{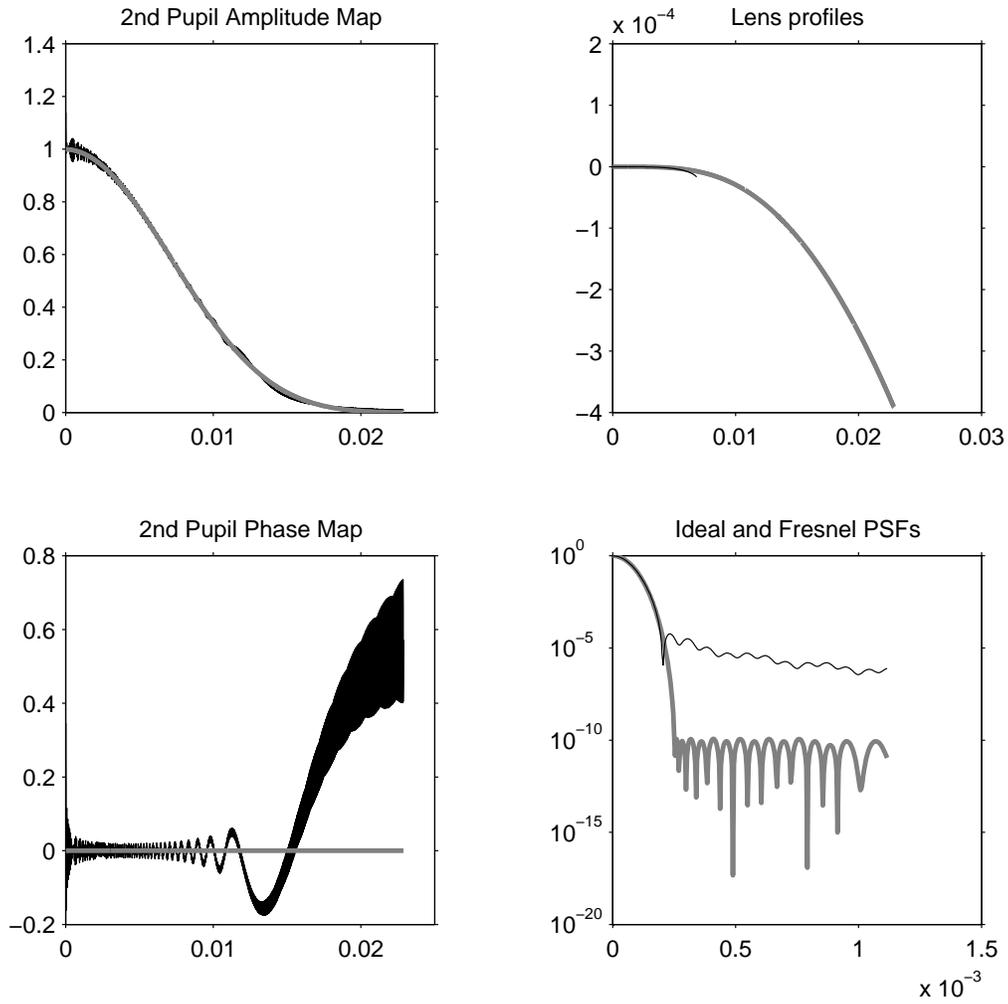}}
\end{center}
\caption{Same as in Figure \ref{fig:7} but with the apodization function
normalized to a maximum value of one.  This normalization results in the two
lenses having different apertures, the second is about 4 times that of the
first.
}
\label{fig:9}
\end{figure}

There is no particular reason to make the two lenses have equal aperture.
By scaling the apodization function, we can easily generate examples with
unequal aperture.  One such experiment we tried was to scale the apodization
function so that its value at the center is one.  This scaling results in the
second lens having almost four times the aperture of the first lens.  The
results for this case are shown in Figure \ref{fig:9}.

\begin{figure}
\begin{center}
\text{\includegraphics[width=6.5in]{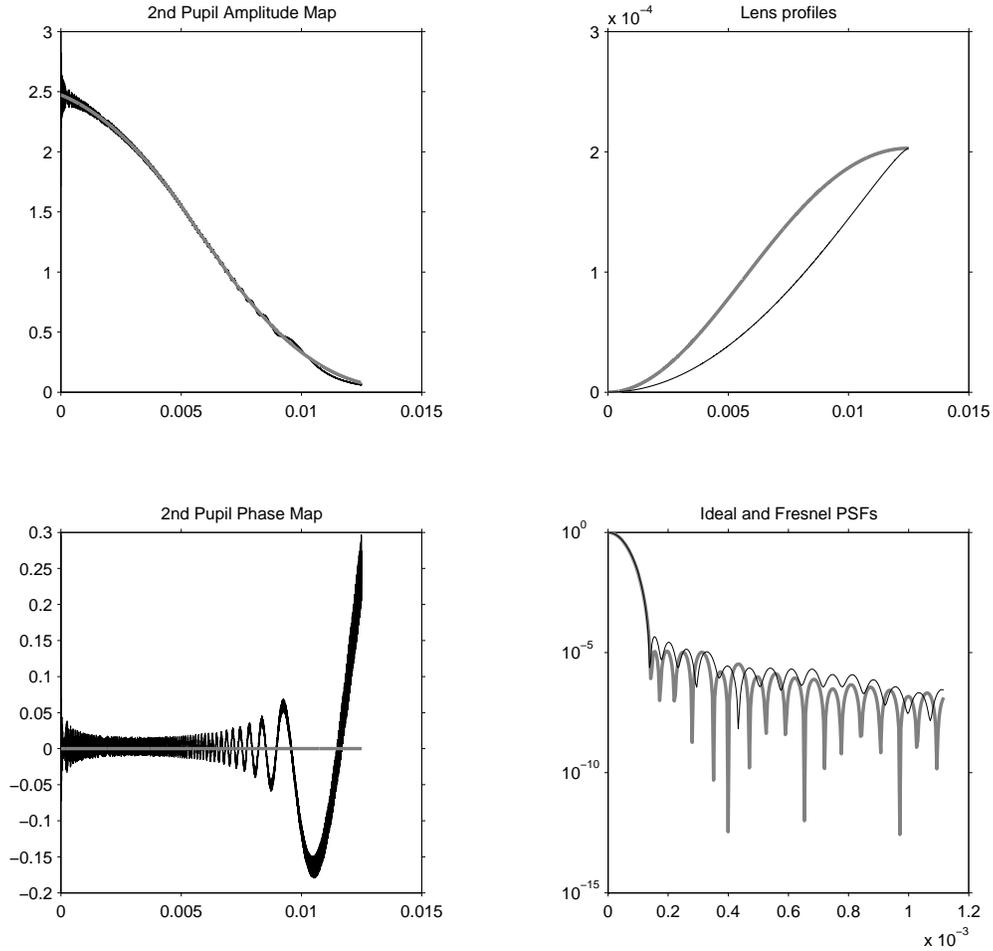}}
\end{center}
\caption{The various sanity checks suggest that diffraction effects 
fundamentally
limit the close-in contrast attainable by pupil mapping to about $10^{-5}$.
So, an apodization profile designed for $10^{-10}$ might not be the
right one to use.  Here, we show results for an apodization profile that
only attempts to achieve $10^{-5}$.  Note that the PSF via
Huygens propagation (black) agrees fairly well with the ideal (Fourier
transform) PSF (gray)
in the lower-right plot.  
}
\label{fig:11}
\end{figure}

If an apodization designed for $10^{-10}$ contrast only produces $10^{-5}$,
one wonders how well an apodization designed for $10^{-5}$ will do.  The
answer is shown in Figure \ref{fig:11}.  In this case, 
the degradation due to diffraction effects is very small.

\section{Hybrid Systems}

\begin{figure}
\begin{center}
\text{\includegraphics[width=6.5in]{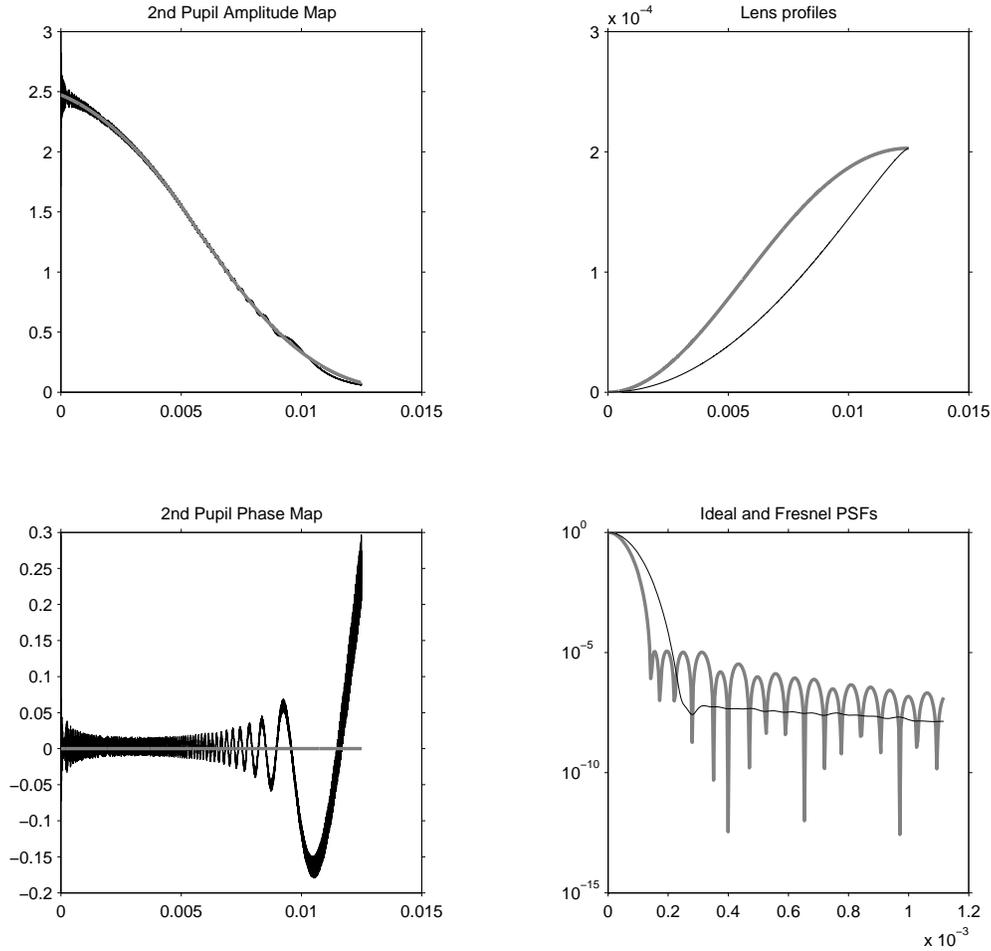}}
\end{center}
\caption{Given our success in Figure \ref{fig:11} of achieving $10^{-5}$
contrast with an apodization profile specifically designed for this level of
contrast, we can now ask whether it is possible to put a conventional
apodizer (or shaped pupil equivalent) in the exit pupil to further ``convert''
this apodization profile into one that achieves $10^{-10}$.  The result is
shown here.  As seen in the PSF plots in the lower right, this combination
gets the contrast to $10^{-7.5}$.
}
\label{fig:12}
\end{figure}

If a pupil mapping system designed for $10^{-5}$ works well, perhaps it could
be followed by a conventional apodizer that attempts to bring the system from
$10^{-5}$ down to $10^{-10}$.  We tried this.  The result is shown in Figure
\ref{fig:12}.  As can be seen in the lower-right plot, the contrast achieved
is limited to about $10^{-7.5}$.  Apparently the diffraction ripples going
into the apodizer are enough to prevent the system from achieving the desired
contrast.

\begin{figure}
\begin{center}
\text{\includegraphics[width=6.5in]{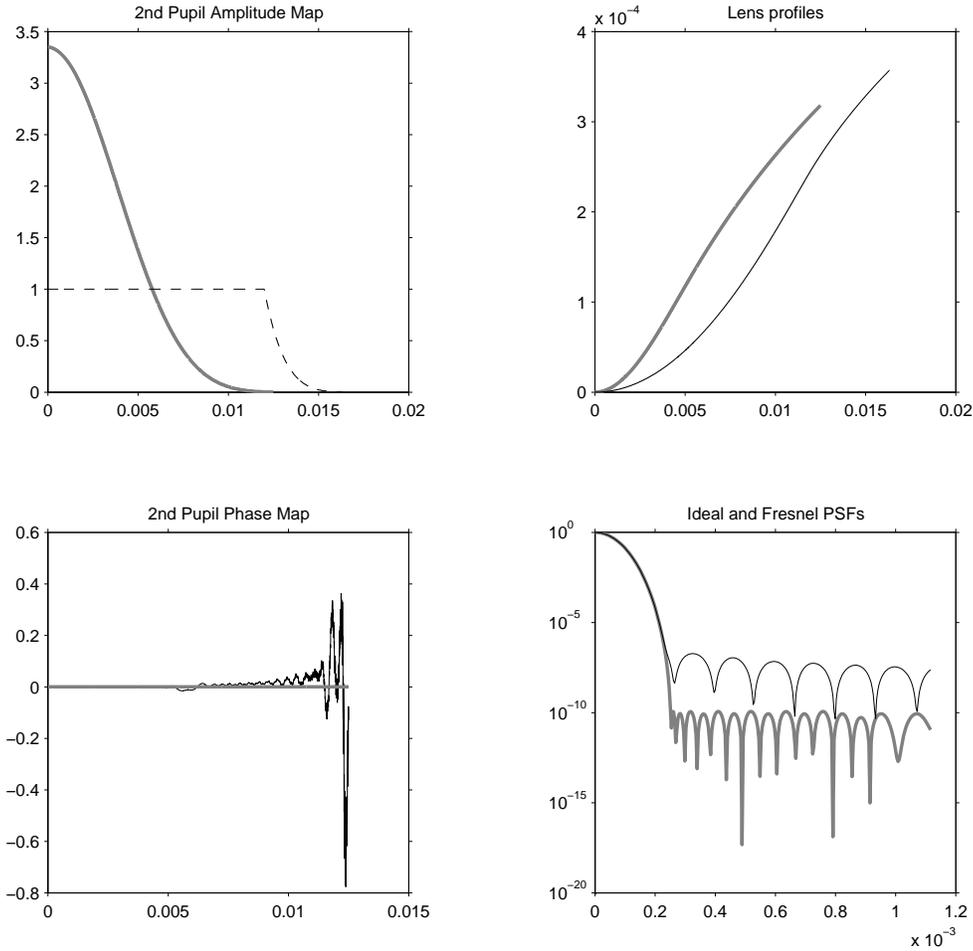}}
\end{center}
\caption{As an alternative to the back-end apodizer considered in Figure
\ref{fig:12}, we consider here the possibility of ``softening'' the edge of
the first lens by using a pre-apodizer.  The green curve in the upper-left
plot shows the pre-apodization function we used.  As one can see from the
lower-right hand plot, this pre-apodization technique allows one to get the
first side-lobe almost down to the $10^{-7}$ level.  This result is not quite
as good as what we had with the post-apodizer of Figure \ref{fig:12} but it is
likely to be more manufacturable.
}
\label{fig:13}
\end{figure}

As an alternative to a downstream apodizer, one could consider a pre-apodizer
placed in front of the first lens.  Since it is generally hard edges that
create bad diffraction effects, we can imagine using the pre-apodizer to
provide the near-outer-edge apodization and allow the pupil mapping system to
provide the main body of the apodization.  In this way, perhaps the
diffraction effects can be minimized while at the same time maintaining a
system with high throughput.  The results of one such experiment are shown in
Figure \ref{fig:13}.

%

{\bf Acknowledgements.}
This research was partially performed for the 
Jet Propulsion Laboratory, California Institute of Technology, 
sponsored by the National Aeronautics and Space Administration as part of
the TPF architecture studies and also under JPL subcontract number 1260535.
The first author received support from the NSF (CCR-0098040) and
the ONR (N00014-98-1-0036).

\bibliography{../lib/refs}   

\begin{thebibliography}{6}
\providecommand{\natexlab}[1]{#1}
\providecommand{\url}[1]{\texttt{#1}}
\expandafter\ifx\csname urlstyle\endcsname\relax
  \providecommand{\doi}[1]{doi: #1}\else
  \providecommand{\doi}{doi: \begingroup \urlstyle{rm}\Url}\fi

\bibitem[Born and Wolf(1999)]{BW99}
M.~Born and E.~Wolf.
\newblock \emph{Principles of Optics}.
\newblock Cambridge University Press, New York, NY, 7th edition, 1999.

\bibitem[Galicher et~al.(2004)Galicher, Guyon, Ridgway, Suto, and
  Otsubo]{GGRSO04}
R.~Galicher, O.~Guyon, S.~Ridgway, H.~Suto, and M.~Otsubo.
\newblock Laboratory demonstration and numerical simulations of the
  phase-induced amplitude apodization.
\newblock In \emph{Proceedings of the 2nd TPF Darwin Conference}, 2004.

\bibitem[Guyon(2003)]{Guy03}
O.~Guyon.
\newblock Phase-induced amplitude apodization of telescope pupils for
  extrasolar terrerstrial planet imaging.
\newblock \emph{Astronomy and Astrophysics}, 404:\penalty0 379--387, 2003.

\bibitem[Traub and Vanderbei(2003)]{TV03}
W.A. Traub and R.J. Vanderbei.
\newblock {Two-Mirror Apodization for High-Contrast Imaging}.
\newblock \emph{Astrophysical Journal}, 599:\penalty0 695--701, 2003.

\bibitem[Vanderbei et~al.(2003)Vanderbei, Spergel, and Kasdin]{VSK03}
R.J. Vanderbei, D.N. Spergel, and N.J. Kasdin.
\newblock {Circularly Symmetric Apodization via Starshaped Masks}.
\newblock \emph{Astrophysical Journal}, 599:\penalty0 686--694, 2003.

\bibitem[Vanderbei and Traub(2005)]{VT04}
R.J. Vanderbei and W.A. Traub.
\newblock {Pupil Mapping in 3-D for High-Contrast Imaging}.
\newblock \emph{Astrophysical Journal}, 2005.
\newblock To appear.

\end{thebibliography}
\bibliographystyle{plainnat}   

\clearpage

\begin{figure}
\begin{center}
\text{\includegraphics[width=6.5in]{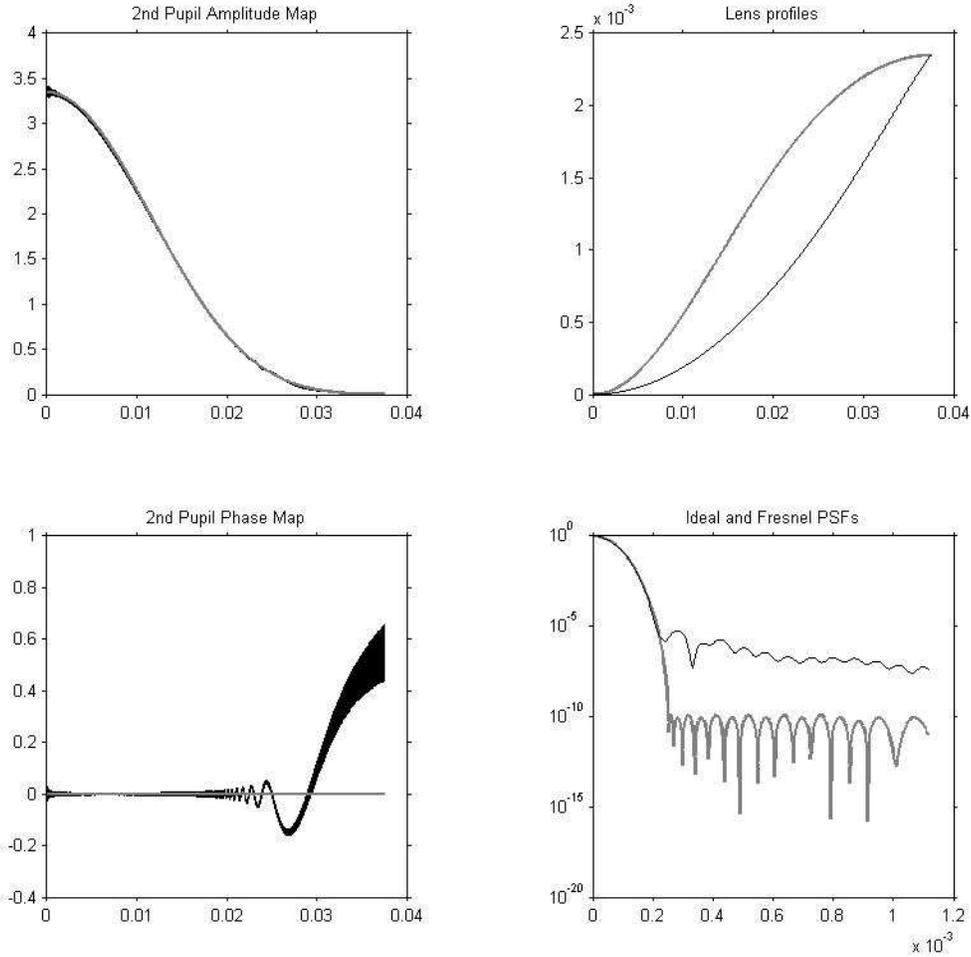}}
\end{center}
\caption{Same as in Figure \ref{fig:7} but with $75$mm lenses ($3$ times
larger).  The larger optical elements required a finer discretization of the
integrals: $30,000$ points were used.  The contrast improves but 
only marginally.
}
\label{fig:10}
\end{figure}

\clearpage

\end{document}